\documentclass[conference]{IEEEtran}
\IEEEoverridecommandlockouts
\usepackage{cite}
\usepackage{graphicx}
\usepackage{tikz}
\usepackage{amsfonts}
\usepackage{amsmath}
\usepackage{algorithm}
\usepackage[noend]{algpseudocode} 
\usepackage{balance}
\usepackage{multirow}
\usepackage[symbol]{footmisc}
\usepackage{bbding}
\usepackage{array}
\usepackage{booktabs}
\usepackage{colortbl}
\usepackage{xcolor}
\usepackage{makecell}
\usepackage{pifont}
\usepackage{xurl}
\usepackage{enumitem}
\usepackage{stackrel,amssymb}
\usepackage[dvipsnames]{xcolor}
\usepackage{hyperref}

\def\BibTeX{{\rm B\kern-.05em{\sc i\kern-.025em b}\kern-.08em
    T\kern-.1667em\lower.7ex\hbox{E}\kern-.125emX}}

\newcommand\ours[0]{{ETC}}
\newcommand\prevstate[0]{\textit{prev state}}
\newcommand\currstate[0]{\textit{curr state}}

\newcommand{\MyPara}[1]{\vspace{.1em}\noindent\textbf{\textit{#1}}~~}

\newcommand{\tabincell}[2]{\begin{tabular}{@{}#1@{}}#2\end{tabular}}

\begin{document}

\title{Direct Model State Migration for Elastic Training of Large Language Models}

\author{
\IEEEauthorblockN{Weijian Liu$^{1,2}$, Mingzhen Li$^{1,2\ast}$, Rui Kang$^3$, Chen Sun$^3$, Guangming Tan$^{1,2}$, Weile Jia$^{1,2\ast}$}
SKLP, Institute of Computing Technology, CAS$^1$ \\
University of Chinese Academy of Sciences$^2$ \\
Huawei Technologies Co., Ltd.$^3$

}

\maketitle

\begin{abstract}

Large language model (LLM) training shall adapt to dynamic resources in shared clusters to tackle the elasticity, including passive preemption and optimistic scaling. 
State migration across device sets is required when altering the hybrid-parallel configuration due to dynamic resources. 
Existing solutions rely on checkpoint-based mechanisms, which persist complete states to storage for resuming with re-assigned resources, forcing all GPUs to stall when transferring model states.
Despite optimization efforts, checkpoint-based solutions incur prohibitive latency (tens to hundreds of seconds) due to data movement across memory hierarchies and disregard for GPU-resident state locality.
We propose ETC, a checkpoint-free state migration framework for elastic hybrid-parallel LLM training. 
We exploit the state locality to minimize inter-GPU data movement, replacing persistence with direct device-to-device communication. 
We leverage the cost matrix abstraction to quantify data movement between parallel configurations and find the optimal migration sketch through a cost-matrix-driven approach. 
Besides, we eliminate node fragmentation through communication coalescing. 
Integrated with Megatron-LM, ETC reduces migration overhead by 2.33$\times$ to 6.37$\times$ compared to checkpoint-based solutions across diverse parallel configurations. By enabling efficient migration, ETC unlocks practical elastic training in production environments.

\end{abstract}

\section{Introduction}
\label{sec:introduction}

Large language models (LLM), based on the transformer architecture, have emerged as foundational models driving diverse applications. However, training these models incurs substantial computational and memory demands. To achieve state-of-the-art performance in quality and sample efficiency~\cite{kaplan2020scalinglawsneurallanguage}, contemporary LLMs are progressively scaled up in model size.
The immense computational demands and prolonged execution times of LLM training necessitate deployment on shared large-scale clusters, where hybrid-parallel training jobs are submitted to the cluster job scheduler.

In shared clusters, the GPU resources visible to a long-running training job do not remain static. Job arrivals, job completions, quota reclaiming, and scheduler-driven reallocation can continuously reshape the device set assigned to a running job. Therefore, elastic hybrid-parallel training should be prepared to migrate model state whenever the available device set changes, rather than assuming a fixed cluster partition throughout the entire training process.
To achieve better resource utilization, a training job should scale in/out when fluctuating GPU resources become available due to cluster load change, and the scheduler should frequently reconcile the GPU allocation among jobs in a proactive manner (i.e., preemption and scaling), especially on shared clusters~\cite{2025-atc-colocating, 2020-osdi-pipeswitch, 2024-eurosys-orion} that co-locates inference jobs and training jobs.
Therefore, state migration---the process of transferring model states when moving a training job from device set $A$ to device set $B$---is essential for hybrid-parallel training jobs, especially those deployed in shared clusters.

\textit{Standard solutions of state migration for a hybrid-parallel training job predominantly rely on the checkpointing mechanism}, which \textit{checkpoints} complete states (parameters, optimizer states, etc.) to persistent storage before migration and then \textit{resumes} from the states after the restart. 
If the parallel configuration (i.e., pipeline/tensor-parallel dimension) should be altered according to dynamic resources, the state should be \textit{resharded} to match the new parallel configuration before resuming. 
Unfortunately, checkpointing and resharding reside on the critical path of the training loop.
Consequently, frequent migrations triggered by elastic training force \textit{all} participating GPUs to stall computation during the lengthy checkpointing, leading to significant resource under-utilization.

\begin{figure}[t]
  \centering
  \includegraphics[width=0.9\linewidth]{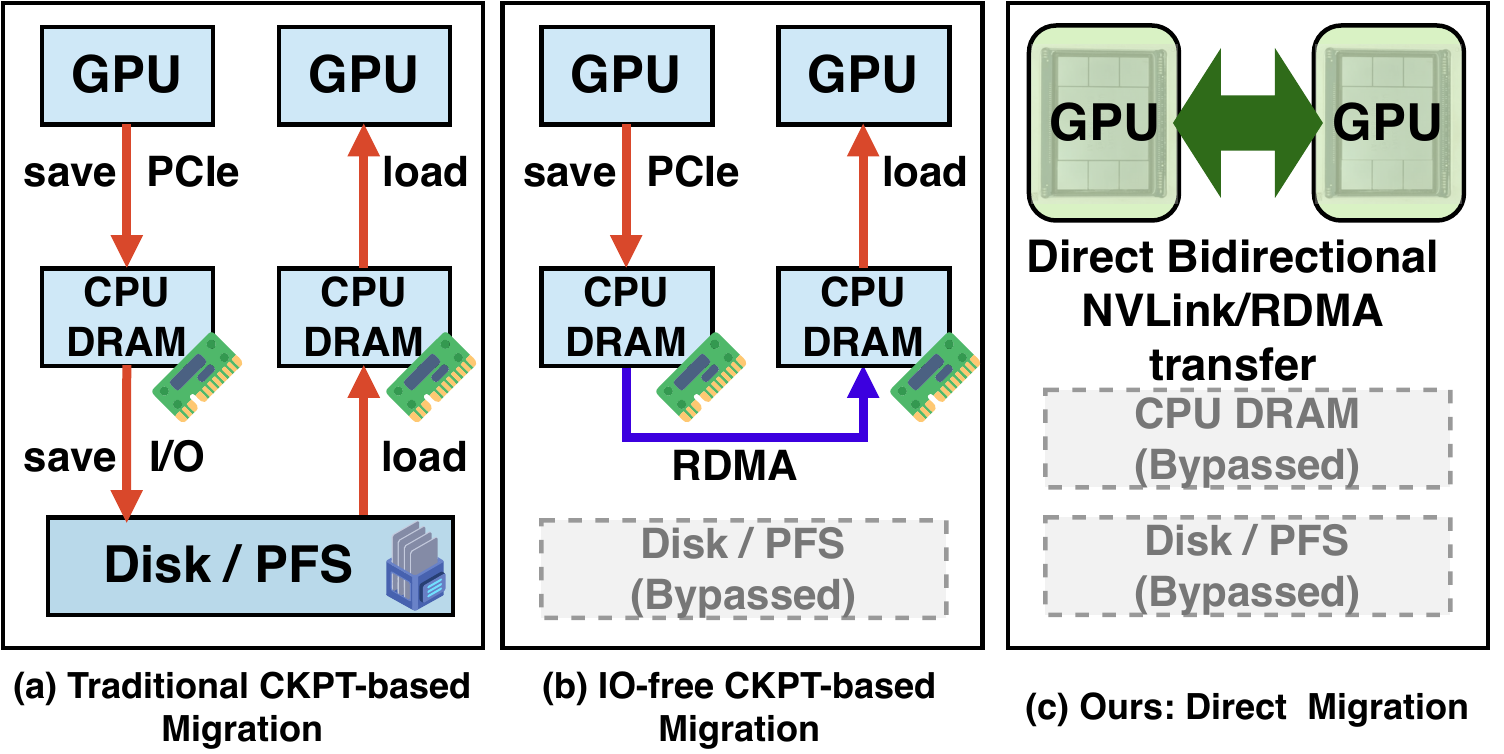}
  \caption{Comparing checkpointing-based migrations and \ours{}.}
  \label{fig:compare_ckpt}
\end{figure}

Recently, many efforts have attempted to reduce checkpointing overhead.
A common practice is to leverage DRAM for intermediate persistence, where the checkpoint is then transferred from DRAM to storage asynchronously~\cite{2021-fast-checkfreq, 2025-asplos-pccheck, dlrover, easyckpt, 2023-sosp-gemini}.
Based on that, 
PCcheck~\cite{2025-asplos-pccheck} splits the checkpoint into chunks to enable concurrent chunk persistence with multiple threads; 
Gemini~\cite{2023-sosp-gemini} presents a placement strategy to maximize failure recovery and a traffic scheduling algorithm to avoid the interference of checkpoint traffic on model training;
Tenplex~\cite{2025-sosp-tenplex} focuses on dynamic parallelism transformation through an in-memory file system and RDMA. 
Besides, Check-N-Run~\cite{2022-nsdi-checknrun} performs differential checkpointing on values and leverages quantization to reduce checkpoint size.
Oobleck~\cite{2023-sosp-oobleck} exploits inherent redundancy of the replicated model states through pipeline templates for fast recovery.
Despite these optimization efforts, checkpoint-based migration remains fundamentally costly. Even the state-of-the-art techniques still incur substantial latency (often tens to hundreds of seconds~\cite{2025-asplos-pccheck, 2025-sosp-tenplex}) because they inevitably involve moving massive amounts of state tensors across the memory hierarchy (e.g., GPU memory$\to$host DRAM$\to$disk). As demonstrated in Figure~\ref{fig:compare_ckpt} ,this process suffers from the bandwidth limitations of storage devices, and \textit{disregards the crucial state locality residing in GPU memory}, necessitating the expensive transfer and persistence of the \textit{entire} model state.

To overcome above issues, we propose \ours{}, a checkpoint-free state migration framework with minimal data movement guarantee for elastic hybrid-parallel LLM training.
\ours{} can migrate the model states of a hybrid-parallel configuration (e.g., data/pipeline/tensor parallel) to the states of another configuration (denoted as reconfiguration).
Unlike other checkpoint-based migrations, 
all migrations in \ours{} are completely through inter-device communication with no need of any heavy I/O persistence, considering the common practice of equipping state-of-the-art GPUs (or NPUs) with high-bandwidth interconnects. 
The key idea of \ours{} is exploiting the model state locality to keep more states in-place, so as to minimize the data movement during migration. 
Note that \ours{} can be integrated with automatic parallel frameworks (e.g., Alpa~\cite{2022-osdi-alpa}, Unity~\cite{2022-osdi-unity}, AutoDDL~\cite{2024-tpds-autoddl}), as \ours{} switches model states according to the timely generated configurations efficiently, which together optimize the holistic process of elastic hybrid-parallel training.

\ours{} provides a collective communication primitive for the state migration of hybrid-parallel reconfiguration.
The primitive takes the current \& previous parallel configuration, and model state as input (\S\ref{sec:design}), and then derives the cost matrix to describe data movement (i.e., quantity of model states requiring transmission) between current ranks and previous ranks (\S\ref{sec:cost_matrix} for pipeline parallel and \S\ref{sec:cost_matrix_extended} for tensor/data parallel) and derive coarse-grained migration sketch, and then generates the fine-grained device-to-device communication instructions for migration (\S\ref{sec:engine}), and then performs resource de-fragmentation through additional migration operators passed to the execution engine (\S\ref{sec:rank_allocator}). 
We also coalesce the migration instructions to avoid the redundant indirect communications.
Specifically, the key contributions are as follows:

\begin{itemize}
\item We formalize a cost matrix to represent the data movement across devices, and employ a cost-matrix-driven approach to minimize the data movement globally.
\item We design the execution engine to transform the coarse-grained migration sketch into fine-grained \texttt{Send}, \texttt{Recv}, and \texttt{Refer} instructions.  
\item We propose a virtual rank allocator, which decouples the virtual and physical process ranks to eliminate resource fragmentation. It guarantees sustained training efficiency, resource occupancy, and minimized data movement.
\item We deploy \ours{} in Megatron-LM to enable the elastic training of hybrid-parallel transformer-based models. Evaluation results show that \ours{} can accelerate the model state migration under various model sizes and parallel configurations.
\end{itemize}

\section{Motivation}
\label{sec:motivation}

\subsection{Hybrid-Parallel LLM Training}
\label{sec:motivation_parallelism}

Efficient LLM training requires hybrid-parallelism---the combination of data, tensor, pipeline, and other parallelisms~\cite{2020-vldb-ddp,2020-sc-zero,2019-arxiv-megatron-lm,2022-kdd-deepspeed,2022-osdi-alpa}. 
A large-scale training job should be broken down into sub-tasks and then assigned to multiple devices for parallel execution, thereby balancing computation, memory, and communication among devices.
Currently, the LLM training jobs are executed in fixed degree of parallelism (DoP), and they follow the fixed parallel configuration during training.
\textit{1)} Data parallel (DP)~\cite{2020-vldb-ddp,2020-sc-zero}: It splits the training data into multiple mini-batches. Each device loads a complete model replica, computes gradients of mini-batch, and synchronizes gradients with Allreduce for model update. 
\textit{2)} Pipeline parallel (PP)~\cite{2021-sc-chimera,deepseekai2024deepseekv3technicalreport}: It partitions the model layers into multiple stages, where each device is responsible for one or more stages. And a mini-batch is split into micro-batches and then processed in a pipeline manner. In PP, each device sends/receives intermediate tensors to/from its neighbor using peer-to-peer (P2P) communications. 
\textit{3)} Tensor parallel (TP)~\cite{2019-arxiv-megatron-lm,2023-megatron3}: It shards the model parameters at tensor level, and places each shard on a device. 
When a full parameter or activation tensor is required, TP uses collective communications to reassemble the full tensor from corresponding shards.

\begin{table}[htbp]
  \centering
  \footnotesize
  \setlength\tabcolsep{2pt}
  \caption{Comparison across different state migration approaches.}
  \begin{tabular}{cccc}
    \toprule
    \textbf{Method} & \textbf{I/O-Free} & \textbf{Path} & \textbf{Data Movement} \\
    \midrule
    Megatron-LM~\cite{2019-arxiv-megatron-lm} & \textcolor{black}{\ding{55}} & GPU$\leftrightarrows$DRAM$\leftrightarrows$Disk & full+full \\
    TorchElastic~\cite{pytorch_elastic} & \textcolor{black}{\ding{55}} & GPU$\leftrightarrows$Disk & full+\ding{55} (on DP)  \\
    PcCheck~\cite{2025-asplos-pccheck} & \textcolor{black}{\ding{55}} & GPU$\leftrightarrows$DRAM$\leftrightarrows$PMEM & full+full \\
    Varuna~\cite{2022-eurosys-varuna} & \textcolor{black}{\ding{55}} & GPU$\leftrightarrows$Disk (on PP) & full+diff \\
    EasyScale~\cite{2023-sc-easyscale} & \textcolor{black}{\ding{55}} & GPU$\leftrightarrows$Disk & full+\ding{55} (on DP) \\
    ByteCkpt~\cite{2025-nsdi-bytecheckpoint} & \textcolor{black}{\ding{55}} & GPU$\leftrightarrows$DRAM$\leftrightarrows$Disk & full+/ \\
    Oobleck~\cite{2023-sosp-oobleck} & \textcolor{OliveGreen}{\ding{51}} & GPU$\leftrightarrows$GPU (on PP) & /+diff \\
    Tenplex~\cite{2025-sosp-tenplex} & \textcolor{OliveGreen}{\ding{51}} & GPU$\leftrightarrows$DRAM & full+diff \\
    ETC (ours) & \textcolor{OliveGreen}{\ding{51}} & \textcolor{OliveGreen}{\textbf{GPU$\leftrightarrows$GPU (always)}} & \textcolor{OliveGreen}{\textbf{/+diff}} \\
    \bottomrule
    \multicolumn{4}{l}{\footnotesize \tabincell{l}{Note: In Data Movement column, \ding{172}+\ding{173} indicates \ding{172} in persistence and \\ \ding{173} in state resharding. ``full'' represents full states, ``diff'' represents \\ differential states, \ding{55} represents no support, and ``/'' represents no need.}}
  \end{tabular}
  \label{tab:method_comparison}
\end{table}

\subsection{Demands of State Migration under Resource Fluctuation}
\label{sec:motivation_dynamic}

\begin{figure}[t]
  \centering
  \includegraphics[width=0.9\linewidth]{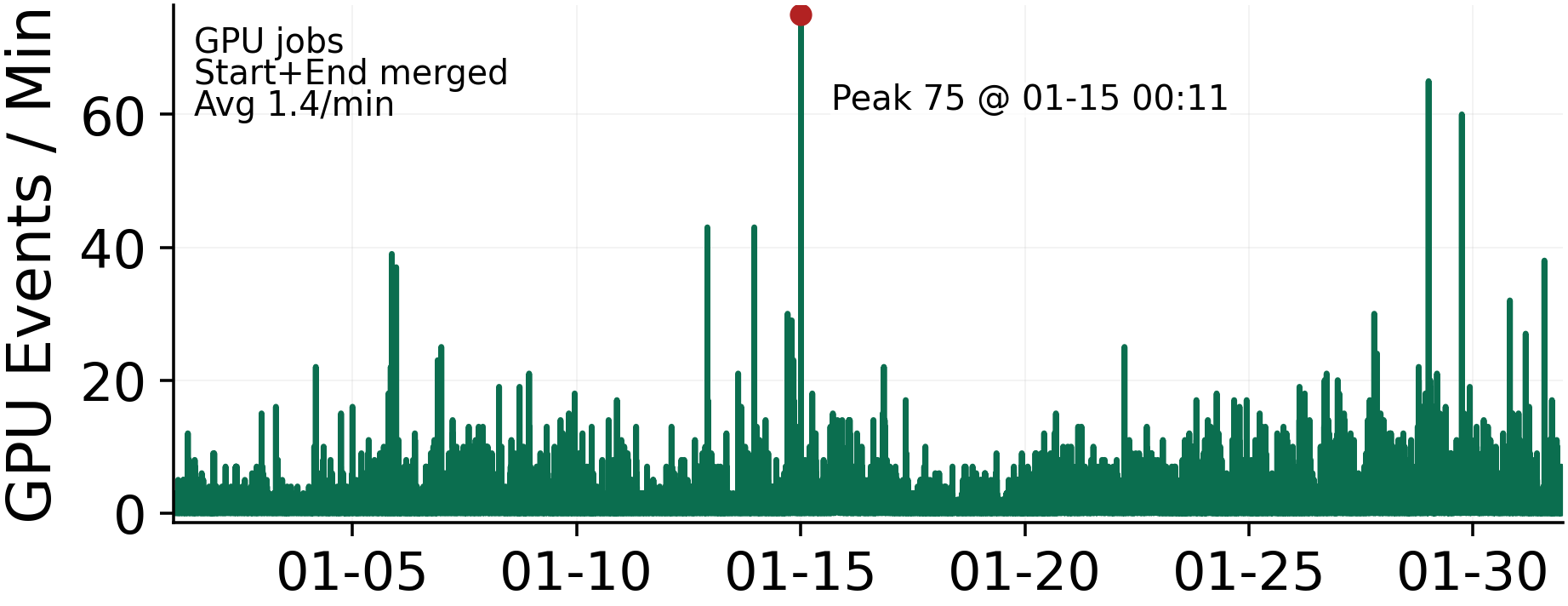}
  \caption{Minute-level GPU-job events of Aurora supercomputer in January 2026. 
  Data is collected from public logs (\url{https://reports.alcf.anl.gov/data/aurora.html }). }

  \label{fig:cluster-volatility-month}
\end{figure}

LLM training relies on large-scale GPU clusters and features a training cycle that spans several days to several months. During this process, the resource fluctuation is common on shared clusters, and dynamic resource changes for jobs are inevitable.
Public traces from Aurora supercomputer show that these disturbances are frequent at the job-event level. In January 2026 alone, the cluster records 32,256/32,249 GPU job starts/ends. 
Figure~\ref{fig:cluster-volatility-month} provides the minute-level start/end events of GPU jobs (i.e., 1.4 events/min on average) and shows that job fluctuation affects long-running training jobs persistently.
Besides, the training/development clusters also exhibit diurnal patterns, and the cluster schedulers usually adopt the minute-level scheduling cycles (e.g., 60s)~\cite{2023-sosp-sia}.
For hybrid-parallel jobs, such fluctuation can force migrations.

These cluster-level fluctuations reach training jobs through several common triggers:
\textbf{\textit{1)} Preemption.} High-priority jobs can take GPU resources from low-priority running jobs~\cite{2022-mlsys-bpt}. For instance, when co-locating inference jobs and training jobs in a shared cluster~\cite{2025-atc-colocating, 2020-osdi-pipeswitch, 2024-eurosys-orion}, training jobs can be preempted by inference jobs to meet strict service-level objectives (SLO)~\cite{2025-asplos-helix, 2025-osdi-nanoflow}. 
Even in training clusters~\cite{2020-osdi-antman, 2023-sc-easyscale}, a training job can be easily preempted due to constant resource revocation. Because the cluster scheduler tends to allocate the currently idle GPUs (but belonging to other users' quota) to the running jobs, when other users submit jobs, these GPUs are reclaimed and the corresponding jobs are preempted.
As for spot instances~\cite{2023-nsdi-bamboo, 2023-sosp-oobleck,2022-eurosys-varuna}, the jobs are more likely to be preempted. For example, a 24-hour trace on AWS EC2 suffers from 127 preemptions~\cite{2023-nsdi-bamboo}. 
\textbf{\textit{2)} Scaling.} To guarantee the strict latency SLO, GPU resources are commonly over-provisioned for inference jobs when tackling bursting inference requests (up to 50$\times$ than the average~\cite{2023-osdi-alpaserve}). It leads to resource under-utilization during off-peak hours~\cite{2025-atc-serverless-in-the-wild,2024-cloud-temposcale,2023-osdi-flux}. 
To saturate GPU resources, training jobs should scale out swiftly to utilize idle GPUs, although they would be preempted or proactively scale in to avoid interference against inference jobs.

\subsection{Opportunity: Minimize Data Movement in State Migration}
\label{sec:motivation_opportunity}
The change of device number must modify the parallel configuration and thus disrupts the collaborative logic among devices, which causes the state migration of training jobs. 
Therefore, to adapt to stage migration, standard solutions usually rely on the checkpointing mechanism (the process of checkpoint--reshard--resume)~\cite{megatron-dist-checkpoint,pytorch-ucp}. 

However, as shown in Table~\ref{tab:method_comparison}, they suffer from huge data movement overhead over migration, hindering their adaptation to frequent-migration scenarios.  
Specifically, 
\textit{1)} Checkpointing enforces the persistence through I/O, including both directly persisting to disk and redirecting to disk from CPU DRAM. In that scenario, the data movement can be the full model state (i.e., a replica of parameters and optimizer states). While the approaches targeting migration within the limited parallel configurations (e.g., Varuna focus on pipeline dimension) should also persist the partial state.
\textit{2)} When changing hybrid-parallel configurations, existing approaches perform state transformation (i.e., reshard) on persisted checkpointing files, leading to additional access to the full model state.
\textit{3)} The state locality is usually ignored because of unawareness of the device ranks. Consequently, they fail to transmit the differential states.
\textbf{Therefore, to enable efficient state migration, we should try to minimize the data movement. We turn to a checkpoint-free approach by keeping model state in GPU memory and performing reconfiguration with inter-GPU connections. }

Closely related to our study is Oobleck~\cite{2023-sosp-oobleck} and Tenplex~\cite{2025-sosp-tenplex}. However, Oobleck only support migration among pipeline templates, while it sacrifices the job's fine-grained control over the parallelism, rendering it \textit{inadequate} for general-purpose migration across arbitrary device sets; and Tenplex employs an in-memory file system for persistence and differential resharding, but still requires transferring full model states with CPU DRAM.

\section{\ours{} Design}
\label{sec:design}

\subsection{Overview}
\label{sec:overview}

\ours{} is a model state migration framework containing the collective communication primitive for various hybrid-parallel training frameworks. 
\ours{} can migrate model states during the hybrid-parallel reconfiguration. 
\textit{\textbf{The design principle of \ours{} is to exploit the locality of model states and eliminate the redundant data movement across devices. }}
Unlike all the other works, \ours{} does not require persistent storage and can directly migrate from one hybrid-parallel state to another, which enables rapid model state migration. 
In addition, \ours{} can retain the original process context and reuse the context, enabling rapid resumption. 

\begin{figure}[t]
    \centering
    \includegraphics[width=\linewidth]{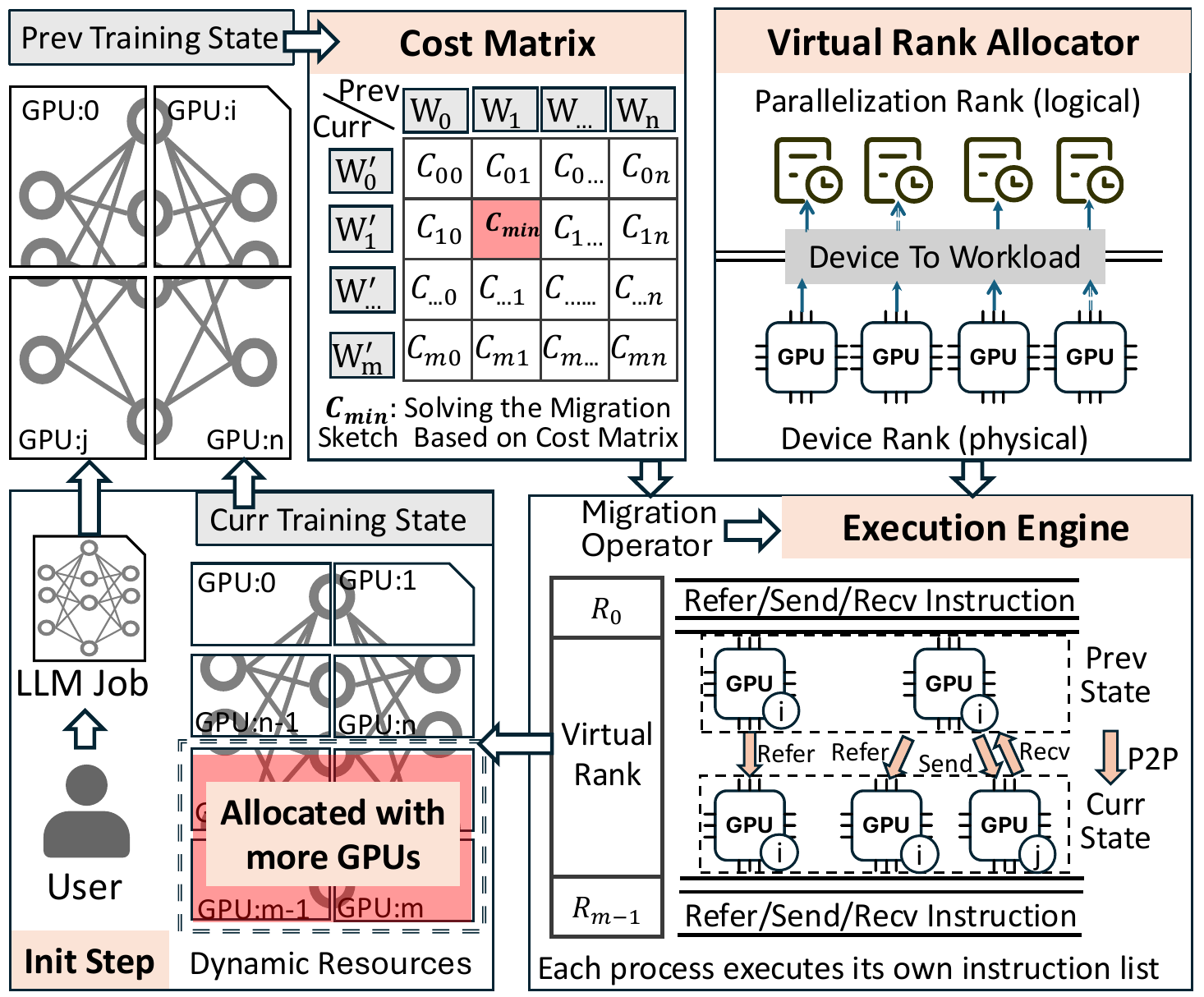}
    \caption{Design overview of \ours{}.}
    \label{fig:overview}
\end{figure}

As illustrated in Figure~\ref{fig:overview}, a user submits a LLM training job to the cluster. Once execution, the job will obtain a hybrid-parallel training state. 
When a migration occurs, this training state is denoted as the \prevstate{}, and \ours{} is notified with the target state after migration, denoted as \currstate{}. Both states are mainly determined by the hybrid-parallel configurations. 
In step 1, we calculate the migration \textbf{cost matrix} between two states. Through the cost matrix, we can obtain the pairwise data movement overhead for any process pair in \prevstate{} and \currstate{}. And we can then derive the optimal \textit{migration sketch} with minimal data movement.
In step 2, we leverage the \textbf{execution engine} to emit the coarse-grained migration operators of the migration sketch. It ultimately executes the direct state migration across GPUs through the fine-grained inter-GPU communication instructions we designed.
In step 3, we employ a \textbf{virtual rank allocator} to provide us more flexibility to adjust the roles of devices. 
It can help perform node de-fragmentation that appears in checkpoint-free migration.
The de-fragmentation is also represented through migration operators and passed to the execution engine. 
Additionally, the execution engine proactively performs communication coalescing to further minimize the data movement.
\ours{} is implemented based on Megatron-LM, and it can be ported to other training frameworks, benefiting from the abstracted workload, migration sketch, virtual rank, and coarse-grained migration operators.

\subsection{Cost-Matrix Driven Migration}
\label{sec:cost_matrix}
\MyPara{Cost Matrix: Quantity the Data Movement between Parallel Configurations.}
During state migration, the states need to be redistributed across devices. 
Specifically, the states are resharded and then transferred from one device to another, where we denote the inter-device communication volume as \textit{data movement}.  
Therefore, we propose the cost matrix to quantify the data movement. 
Cost matrix is derived from the difference of workloads among all process pairs \textit{$\langle$a,b$\rangle$}, where $a$ / $b$ represents the process rank of the hybrid-parallel configuration before / after migration.

\begin{figure}[htbp]
    \centering
    \includegraphics[width=0.85\linewidth]{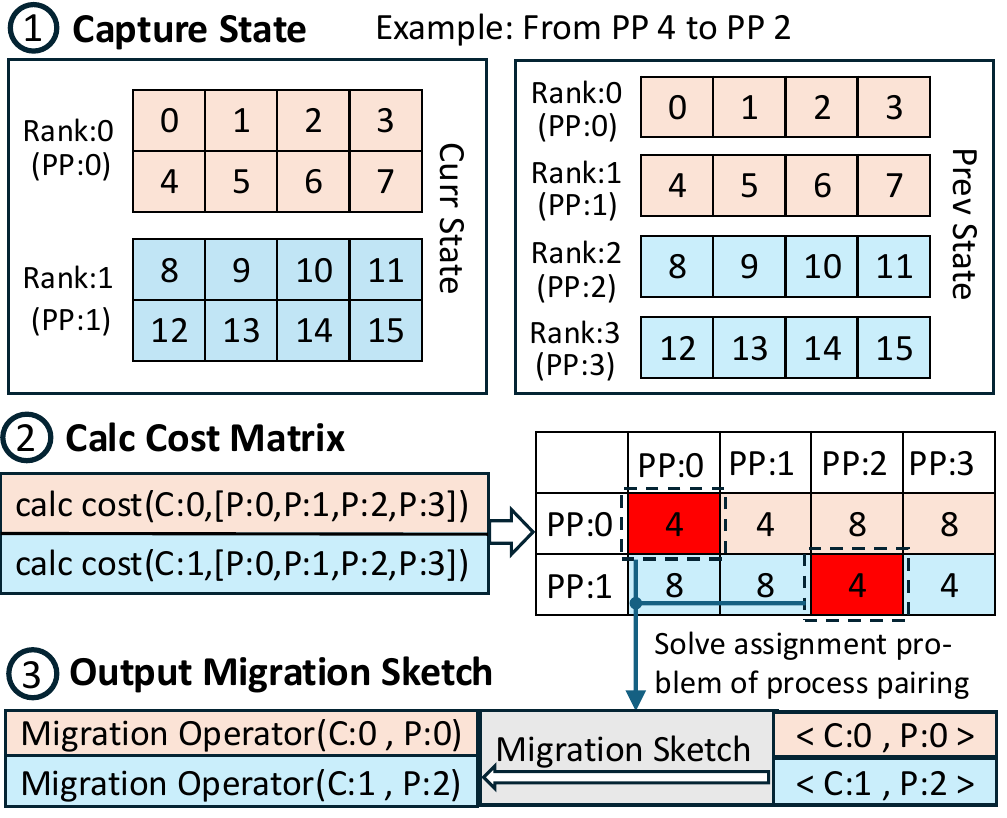}
    \caption{Cost-matrix driven migration. An example of reconfiguring the PP size from 4 to 2 of 16 transformer layers. }
    \label{fig:Matrix-Driven}
\end{figure}

As shown in Figure~\ref{fig:Matrix-Driven}, we first capture the state information (including \prevstate{} and \currstate{}) residing on each process (\ding{172}).
According to the captured states, we express the logical \textit{workload} of the processes in both states. Note that workload is the abstracted representation of tensors, and each item in workload represents a tensor in each process (e.g., \textit{[0]} represents the first transformer layer, and \textit{[0.a]} represents the first TP slice of the first transformer layer). 
Then we build the cost matrix $C\in\mathbb{Z}^{M\times N}$, where the row indices represent the ranks in the \currstate{} ($M$ ranks in total), and the column indices represent the ranks in the \prevstate{} ($N$ ranks in total) (\ding{173}). 
Note that in hybrid-parallel scenarios, the ranks can also be expressed as tuple of ranks in all parallel dimensions (e.g., \texttt{P0T1} represents rank 0 in PP and rank 1 in TP).
Each element $c_{i,j}$ in the cost matrix is derived through $ c_{i,j} = len(curr_i) - len(curr_i \cap prev_j) $ , where $curr_i$ represents the workload list in \currstate{} and $prev_j$ represents the workload list in \prevstate{}. 
Therefore, $c_{i,j}$ can quantify the data movement required to transform previous rank $j$ to current rank $i$.

We provide a example of reconfiguring a model (containing 16 transformer layers) from PP4 to PP2 in Figure~\ref{fig:Matrix-Driven}. 
At \prevstate{}, there are 4 devices participating, where each holds 4 transformer layers, the \textit{workload} of rank 0 is \textit{[0,1,2,3]}. 
When migration, only 2 devices are needed to continue the training. 
In this scenario, each device should hold 8 transformer layers and the \textit{workload} of rank 0 is \textit{[0,1,2,3,4,5,6,7]}. 
The $c_{0,0}$ can be derived by comparing the workloads, and its value is $4$.  

Note that the cost matrix should be synchronized across all devices, so that each device can recognize its responsibility and perform migration in a decentralized manner. Besides, in hybrid-parallel training frameworks that follow the static model state partition rules (e.g., Megatron-LM partitions tensors evenly in PP and TP dimension), each device can derive its workload following these rules and calculate global cost matrix locally, which helps to avoid synchronization overhead.

\MyPara{Migration Sketch: Pairing Current and Previous Processes through Cost Matrix.}
By exploiting the cost matrix, we pair the processes in \prevstate{} with the processes in \currstate{}, so as to maximize the reuse of the previous process context and thus minimize the data movement. 
The \textit{migration sketch} is a series of migration operators on process pairs (e.g., \textit{$\langle$a,b$\rangle$}), which can be translated into fine-grained migration instructions and executed by the execution engine (\S\ref{sec:engine}).
Note that the migration operator for process pair \textit{$\langle$a,b$\rangle$} should transform process $b$ of \prevstate{} into process $a$ in \currstate{}.
The process pairing can be formulated as an assignment problem. 
$c_{ij}$ denotes the cost when the $i$-th process in \currstate{} is associated with the $j$-th process in \prevstate{}. 
We define the indicator variable $x_{ij}$ as:
\vspace{-0.1in}
\begin{equation}
x_{ij}=
\begin{cases}
1, & \text{if the } i\text{-th rank is paired to the } j\text{-th rank} \\
0, & \text{otherwise}
\end{cases}
\end{equation}
\vspace{-0.1in}

We try to minimize the total data movement among all process pairs, so that the objective function is as follows:

\vspace{-0.1in}
\begin{equation}
\label{equ:objective}
\min \quad \sum_{i=1}^{m} \sum_{j=1}^{n} c_{ij} x_{ij}
\end{equation}
\vspace{-0.1in}

When the process number of \currstate{} is larger than that of \prevstate{} (i.e., $M > N$), there are two constraints (Equation~\ref{equ:constraint-1}).
\textit{a)} Each process in the \prevstate{} must be paired with one process in the \currstate{}; and \textit{b)} Each process in \currstate{} can be paired with at most one process in \prevstate{}.

\vspace{-0.1in}
\begin{equation}
\label{equ:constraint-1}
\begin{cases} 
\sum_{i=1}^{m} x_{ij} = 1, & \forall j = 1,2,\dots,n \quad \triangleright\text{Constraint \textit{a)}} \\
\sum_{j=1}^{n} x_{ij} \leq 1, & \forall i = 1,2,\dots,m \quad \triangleright\text{Constraint \textit{b)}} 
\end{cases}
\end{equation}

When the process number of \currstate{} is not larger than that of \prevstate{} (i.e., $M \leq N$), there are two constraints (Equation~\ref{equ:constraint-2}).
\textit{a)} Each process in \prevstate{} must be paired with at most one process in \currstate{}; and \textit{b)} Each process in \currstate{} must be paired with one process in \prevstate{}.

\vspace{-0.2in}
\begin{equation}
\label{equ:constraint-2}
\begin{cases} 
\sum_{i=1}^{m} x_{ij} \leq 1, & \forall j = 1,2,\dots,n \quad \triangleright\text{Constraint \textit{a)}} \\
\sum_{j=1}^{n} x_{ij} = 1, & \forall i = 1,2,\dots,m \quad \triangleright\text{Constraint \textit{b)}} 
\end{cases}
\end{equation}
\vspace{-0.1in}

The assignment problem of process pairing can be efficiently solved using a non-standard Hungarian algorithm~\cite{2016-7738348}, as illustrated in Algorithm~\ref{algo:assignment}. 
The input to the algorithm is the cost matrix $C\in\mathbb{Z}^{M\times N}$, and the output is the migration sketch (i.e., a series of process pairs). 
In the hybrid-parallel reconfiguration scenarios, $M$ and $N$ are usually not equal. 
To adopt the Hungarian algorithm, the cost matrix is padded to a square matrix with zeros (line~\ref{line:padding}). 
Then, we call the Hungarian algorithm in the SciPy\footnote{Hungarian algorithm in SciPy package. \url{https://docs.scipy.org/doc/scipy/reference/generated/scipy.optimize.linear_sum_assignment.html}} package to obtain the process pairs (line~\ref{line:solving}). 
Its time complexity is $O(k^3)$ for a $k \times k$ cost matrix. However, in our cost matrix, $k$ is usually less than 1,024, because \ours{} just need to cover the PP dimension ($\leq 16$) and TP dimension ($\leq 16$), and thus the solving time is just tens of milliseconds on a standard server (refer to \S\ref{sec:cost_matrix_extended} for details). Note that the solving time is much less than a training iteration.
Finally, we remove the processes involved in the padding part from the process pairs (line~\ref{line:remove_start}--\ref{line:remove_end}). 
The output (process pairs) is the optimal migration sketch.

\begin{algorithm}[H]
\caption{Generate migration sketch via process pairing. }
\footnotesize
\label{algo:assignment}
\begin{algorithmic}[1]
\Require cost matrix: $C\in\mathbb{Z}^{M\times N}$
\Ensure migration sketch: \textit{sketch} 
\Function{solve\_cost\_matrix}{$C$}
\State $M, N \gets rows(C), cols(C)$, $k \gets \max(M,N)$
\State \textbf{Zero-padding} matrix $C$ to the size of $k \times k$ \label{line:padding}
\State \# Call Hungarian algorithm 
\State \textit{pairs} $\gets$ \textbf{Hungarian}($C$) \Comment{\textit{Solving}} \label{line:solving}
\For{ \textit{pair $\langle$i,j$\rangle$} in \textit{pairs} } \label{line:remove_start}
    \If{i $\geq$ M or  j $\geq$ N}
        \State Drop \textit{pair $\langle$i,j$\rangle$} from \textit{pairs}
    \EndIf \label{line:remove_end}
\EndFor
\State \textit{sketch} $\gets$ \textit{pairs} \Comment{\textit{Derive the migration sketch}}
\State \textbf{return} \textit{sketch} 
\EndFunction
\end{algorithmic}
\end{algorithm}

\subsection{Extend Cost Matrix to Hybrid-Parallel}
\label{sec:cost_matrix_extended}

\MyPara{Workload Abstraction and Cost Matrix of TP.}
A one-dimensional (e.g., PP) workload can be represented by a single integer (e.g., \textit{[0]}), and a multi-dimensional (e.g., PP + TP) workload should describe the logical workloads of multiple dimensions in a similar manner. 
Unlike PP, TP does not have an inherent layer structure for partitioning and indexing. 
Therefore, we further extend the \textit{workload} for TP.
We use the least common multiple (LCM) of TP size in \prevstate{} and TP size in \currstate{} to determine the total workload in TP dimension, so that the workload of a TP rank (e.g., rank 0 in a TP=4 group) can be represented by a single alphabet (e.g., \textit{[a]} of \textit{[a,b,c,d]}). 
Then the cost matrix of TP can be derived by differentiating the workloads in \prevstate{} and \currstate{} through $ c_{i,j} = len(curr_i) - len(curr_i \cap prev_j) $, which is similar to PP (refer to \S\ref{sec:cost_matrix}).

\begin{figure}[htbp]
    \centering
    \includegraphics[width=\linewidth]{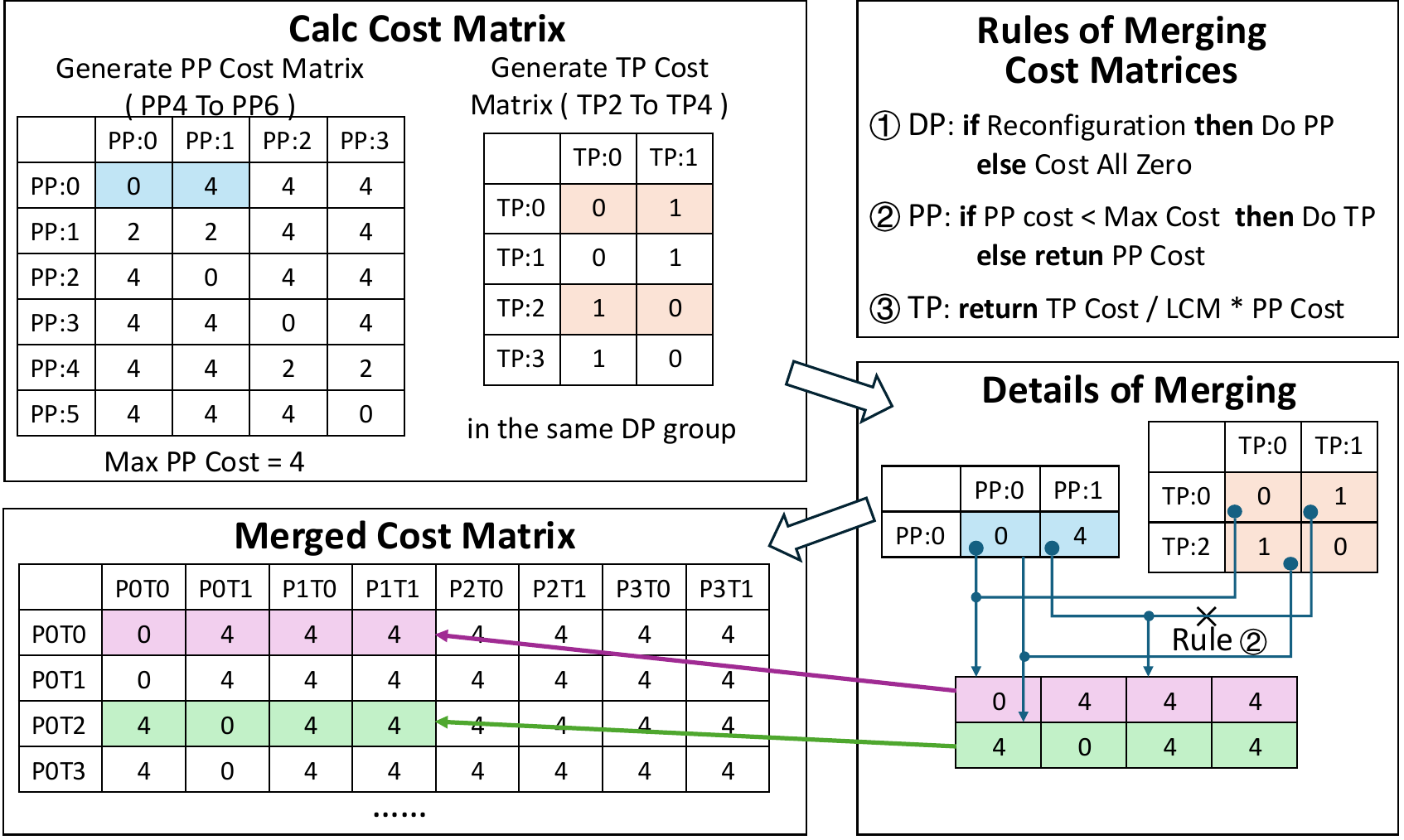}
    \caption{Merging PP cost matrix and TP cost matrix. An example: PP changes from 4 to 6 and TP changes from 2 to 4 simultaneously.}
    \vspace{-0.2in}
    \label{fig:Cost_Matrix_Tensor}
\end{figure}

\MyPara{Merge the Cost Matrices of PP and TP.}
Under the PP+TP scenario, as a process has individual ranks in PP group and TP group (e.g., \texttt{P0T0}), the workload is represented by the nested PP workload and TP workload (e.g., \textit{[0.a, 1.a]}).
Once we derive the individual cost matrices of PP and TP, we should merge them into a monolithic cost matrix. 
For each cost matrix, there is a maximum cost value in the process pairs where the two workloads have no intersection at all. 
Two processes can intersect in the TP dimension only if they also intersect in the PP dimension; equivalently, no TP intersection implies no PP intersection.
For brevity, we set the priority of PP to be higher than that of TP during the merging. 
We can merge cost values dimension by dimension.
\textit{1)} If the cost of the high-priority dimension equals the maximum value of the cost matrix for that dimension (denoted as \textit{max}), it indicates that there is no intersection at all between the process pair, so that the merged cost is equal to \textit{max}. 
\textit{2)} If we reach the cost matrix of the lowest-priority dimension, a coefficient is returned, which equals the cost divided by the maximum cost of this dimension. This coefficient is then passed back dimension by dimension for multiplication.
An example is illustrated in Figure~\ref{fig:Cost_Matrix_Tensor}.

\MyPara{DP Dimension.} 
There is no need to capture DP in the merged cost matrix.
Different from PP/TP, DP replicates the model and the replicas are either kept or discarded during reconfiguration, which has no data movement. 
Besides, we isolate the PP/TP processes under different DP ranks.
If two processes have different DP ranks, even if their workloads on PP/TP are the same, we regard the difference between the workloads to be infinite.
Because there is inherent load balance among different DP replicas, where the migration of PP/TP processes is completely identical. If we create data movement across DP replicas, the load balance tends to be broken.
When increasing DP size during reconfiguration, we insert a series of process-to-process migration operators to the migration sketch, where each PP/TP process in existing DP groups sends all model states to its counterpart in new DP groups. And the source groups are selected in a round-robin manner to ensure the load balance among existing DP groups.

\MyPara{Fault-tolerance through DP.}
Since each DP group contains a complete and synchronized model state, DP groups can be regarded as replicas of each other for fault tolerance.
When the dimension of DP is $d$, \ours{} can accommodate at least $d-1$ hardware failures simultaneously.
When a certain device fails, the workload on the failed device can be obtained from other DP groups via the the PP/TP ranks. 
And we append the corresponding migration operator to the migration sketch.

\MyPara{Hybrid-Parallel Migration Sketch.}
After obtaining the multi-dimensional cost matrix, we can derive the migration sketch, the solving procedure is similar to that in \S\ref{sec:cost_matrix}. 
Unlike the one-dimensional migration sketch, the cost matrix index of the hybrid-parallel migration sketch is combinations of multiple dimensions (e.g., \texttt{P0T0}). Meanwhile, the migration sketch within existing DP group is identical.

\subsection{Execution Engine of Migration Sketches}
\label{sec:engine}

The execution engine is responsible for executing the migration sketch generated above. 
Basically, it transforms the coarse-grained migration operators in the sketch into fine-grained migration instructions, and then executes the instructions for the final migration.

\begin{figure}[htbp]
    \centering
    \includegraphics[width=\linewidth]{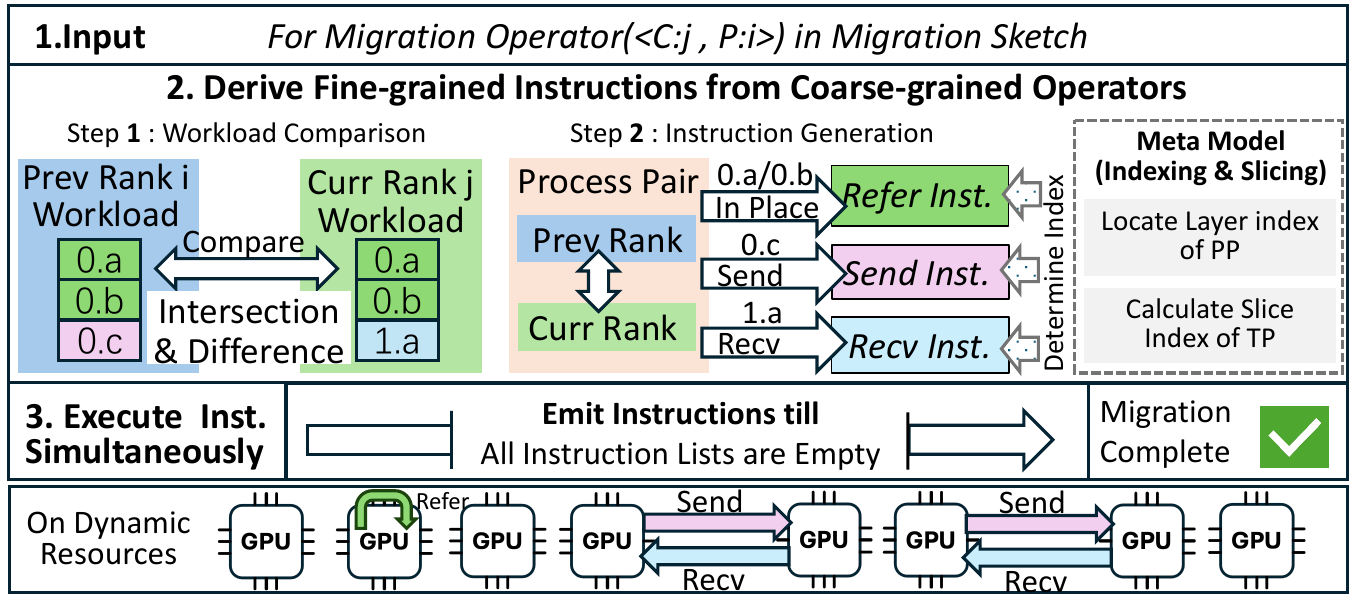}
    \caption{Execution Engine of \ours{}.}
    \label{fig:exec_engine}
\end{figure}

\MyPara{Migration Instructions.}
We design three migration instructions including \texttt{Send}, \texttt{Recv}, and \texttt{Refer}. Each instruction takes the source process id \textit{sid}, destination process id \textit{did}, workload type \textit{wt}, and workload id \textit{wid} as the input. 
Each process pair in the migration sketch can be transformed into one or more instructions. Specifically, as for a process pair \textit{$\langle$curr, prev$\rangle$}, 
\textit{1)} if a workload is not in \textit{curr} process but in \prevstate{}, it will be packed into a \texttt{Send} instruction; 
\textit{2)} if a workload is in \textit{curr} process but not in \prevstate{}, it will be packed into a \texttt{Recv} instruction; 
\textit{3)} if a workload is both in \textit{curr} process and in \textit{prev} process, it will be packed into \texttt{Refer} instruction and stay in place. Note that \texttt{Refer} only binds the workload to the new state, requiring no data movement.

It is necessary to migrate model parameters, optimizer states and other states. 
Therefore, we provide the workload type \textit{wt} variable in migration instructions, and maintain the mapping relationship between the logical workload to different workload types, so that the instructions are compatible to migrate these workload types. The states can be packed to the instructions when \texttt{Send} and then unpacked to update the corresponding state after migration.

\MyPara{Meta Model for Indexing TP Slices.}
Given a nested workload (e.g., \textit{[0.a]}), we can locate the transformer layer with the PP index (e.g., $0$), but we cannot locate the TP slices, because we lack the information that which dimension of the high-dimension tensor is sliced by TP.
Therefore, we initialize a meta model in PyTorch's meta device, which maintains only the tensor shapes without actually loading the actual tensors into memory. 
The TP size of the meta model is set to 1. 
By differentiating the tensor shapes of the meta model and that in \prevstate{}, we can help identifying how the TP dimension is partitioned. 
For example, if there is a tensor dimension satisfying \textit{meta\_model\_dim / prev\_model\_dim == prev\_TP\_size}, this dimension is sliced, and the slice shape corresponding to the logical workload can calculated as \textit{meta\_model\_dim / LCM(curr\_TP\_size, prev\_TP\_size)}.

\MyPara{Execution Flow.}
The complete execution flow is illustrated in Figure~\ref{fig:exec_engine}. 
For each coarse-grained migration operator \textit{$\langle$curr:j, prev:i$\rangle$} in the migration sketch, the execution engine compares the logical workloads of process $i$ in the \prevstate{} and process $j$ in the \currstate{}. 
Then the migration instructions including \texttt{Send}, \texttt{Recv}, and \texttt{Refer} are generated with above three rules \textit{1)2)3)}. 
Then the device derives the index of the tensor slice corresponding to the workload based on the meta model and the workload type $wt$.
At this time, the migration instructions are recorded into an \textit{instruction list} on each device.
With all migration instructions in the migration sketch are ready, all devices emit their instruction lists simultaneously till empty. 

To prevent \prevstate{}\textit{s} and \currstate{}\textit{s} from coexisting, we use \textit{state chunking} to migrate states in sequential chunks with on-demand buffer allocation, and \textit{eager release of source-side state} to promptly free source-side tensors after each round.

\subsection{Virtual Rank Allocator for Resource De-fragmentation}
\label{sec:rank_allocator}
When performing hybrid-parallel reconfiguration with \ours{}, we find the issue of resource fragmentation.
Because \ours{} reuses the process contexts for better locality, the devices are not restarted with a new placement but only assigned with new ranks.
There are two categories of resource fragmentation.

\begin{figure*}[t]
  \centering
  \vspace{-0.1in}
  \includegraphics[width=0.9\linewidth]{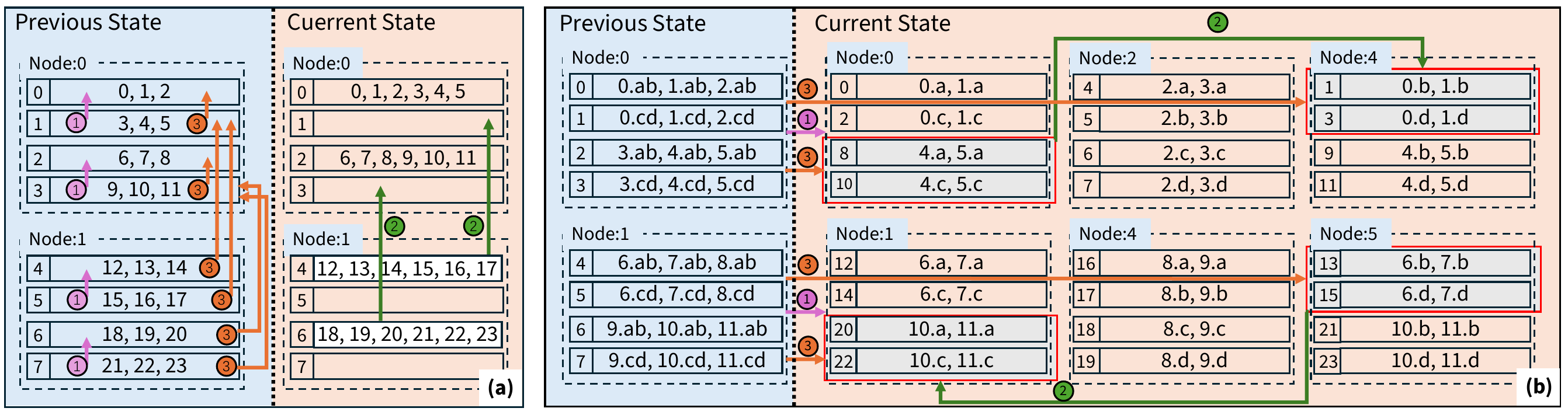}
  \caption{De-fragmentation and coalescing. (a) Occupancy-level. (b) Efficiency-level.
  The arrows \ding{172}\ding{173}\ding{174} represent communication of state migration, de-fragmentation, and their coalescence, respectively.}
  \label{fig:combined_fragmentation}
\end{figure*}

\textit{1) Occupancy-level.}
As shown in Figure~\ref{fig:combined_fragmentation}(a), when scaling in the PP size,  some devices are occupied while others stay idle. From the cluster scheduler's perspective, incoming jobs are infeasible to be assigned to nodes with good occupancy.

\textit{2) Efficiency-level.} 
As shown in Figure~\ref{fig:combined_fragmentation}(b), when the PP size and TP size of a job change simultaneously or the TP size scales out, the cross-node TP groups may arise. This leads to inefficiency in collective communications. 

To avoid resource fragmentation, we propose \textit{virtual rank allocator}.
It isolates the virtual rank within the hybrid-parallel grid and the physical rank within the allocated devices, and manages the mapping between virtual ranks and physical ranks.
Through remapping physical devices with new virtual ranks, it can directly change the roles of physical devices. The remapping can also be represented as migration operators.

\MyPara{Placement-Aware Remapping.}
The key idea to resolve node fragmentation is to adjust the workload on the fragmented nodes through placement-aware remapping. 
The details are shown in Algorithm~\ref{alg:de-fragmentation}.

For occupancy-level de-fragmentation, the remapping is illustrated by  \textit{defragmentation\_for\_occupancy} function. 
Its input includes the world size of \prevstate{} $pw$, the world size of \currstate{} $cw$, and the pairing relationship $p2c$ between the previous processes and the current processes generated from the cost matrix. And the device number in a node is $n$. 
First, we obtain the placement of processes, $pNodes$, $cNodes$, in both \prevstate{} and \currstate{} (line~\ref{line:pp_placement_prev}--\ref{line:pp_placement_curr}).
If a process is dropped after migration, its $id$ is set to -1. All indices in $curr\_nodes$ corresponding to -1 represent idle devices (line~\ref{line:pp_placement_curr}), and the rest are occupied devices. 
We pair the idle devices in order with the occupied device in reverse order, continuing this process sequentially for all remaining devices (line~\ref{line:pp_fragement_exec_start}--\ref{line:pp_fragement_exec_end}), so that we can populate the migration operators through pairing.

For efficiency-level de-fragmentation, the remapping is illustrated by  \textit{defragmentation\_for\_efficiency} function. 
In common practice, the TP group should be placed inside a node for optimal collective communication. 
First, we iterate through all process ids in the \currstate{}, which is the \textit{target} placement recording the optimal placement on physical devices (line~\ref{line:device_nodes}). 
The $loads$ records the placement after state migration (line~\ref{line:load_nodes}), which contains the efficiency-level fragmentation. 
The inconsistent node pairs of $target$ and $loads$ are marked as fragmentation (line~\ref{line:tp_frag_nodes}). 
The process pairs are generated though the processes in fragmentation nodes, which are then passed to the migration operators (line~\ref{line:tp_fragement_exec_start}--\ref{line:tp_fragement_exec_end}).

\begin{algorithm}[htbp]
    \caption{Placement-Aware Remapping.} 
    \label{alg:de-fragmentation}
    \footnotesize
    \begin{algorithmic}[1]
      \Require prev\_world\_size: $pw$, curr\_world\_size: $cw$, pairs: $p2c$, nDevice: $n$ 
      \Ensure migration operators: $mo$
      \Function{defragmentation\_for\_occupancy}{$pw$, $cw$, $p2c$, $n$}
        \State \# Record the process placement
        \State pNodes = [[range(i, i+$n$)] for i in range(1,$pw$,$n$)] \label{line:pp_placement_prev}
        \State cNodes = [[p2c.get(id, -1) for id in node] for node in pNodes]\label{line:pp_placement_curr}

        \State \# Find the indices of idle and occupied devices \label{line:pp_fragement_exec_start}
        \State iter\_idle = iter([i for i, x in enumerate(cNodes) if x == -1])
        \State iter\_occu = iter([i for i, x in enumerate(cNodes) if x != -1]).reversed()
        \While{iter\_idle and iter\_occu are not meeted}
            \State $mo$.append($\langle$iter\_idle.next(), iter\_occu.next()$\rangle$)
        \EndWhile \label{line:pp_fragement_exec_end}
    \EndFunction

    \Function{defragmentation\_for\_efficiency}{$pw$, $cw$, $p2c$, $n$}
        \State \# Generate optimal process placement
        \State target = [[range(i, i+$n$)] for i in range(1,$cw$,$n$)] \label{line:device_nodes}
        \State \# Record current process placement
        \State devices = [[range(i, i+n)] for i in range(1, pw, n)]
        \State loads = [[p2c.get(id) for id in node] for node in devices] \label{line:load_nodes}
        \State frag\_nodes = [(a, b) for a,b in zip(targets, loads) if a != b]\label{line:tp_frag_nodes}
        \For{[pNode, cNode] in frag\_nodes} \label{line:tp_fragement_exec_start}
        \For{pProc,cProc in zip(pNode, cNode)}
        \If{pProc != cProc}
            \State $mo$.append($\langle$pProc,cProc$\rangle$) 
            \EndIf
        \EndFor
        \EndFor \label{line:tp_fragement_exec_end}
    \EndFunction
    \end{algorithmic}
\end{algorithm}

\MyPara{Communication Coalescing.}
To eliminate the redundant data movement, the migration operators of state migration (\S\ref{sec:cost_matrix} and \S\ref{sec:cost_matrix_extended}) and de-fragmentation can be coalesced when the execution engine generates instructions. 
As shown in Figure~\ref{fig:combined_fragmentation}, the state migration is marked by arrow \ding{172}, the de-fragmentation is marked by arrow \ding{173}. 
If \ding{172} sends the workload $w$ from device $i$ to device $k$, and \ding{172} sends the workload $w$ from process $k$ to process $j$, then the execution engine coalesces these two instructions into sending the workload $w$ from process $i$ to process $j$. 
The coalesced instructions are marked by arrow \ding{174}. 
The \texttt{Recv} instructions are coalesced in the same way.

\section{Evaluation}
\label{sec:evaluation}

\subsection{Experimental Setup}
\MyPara{Clusters.} 
We use two different clusters for evaluation. 
\textit{a)} Cluster-A is equipped with 32 nodes connected with $4\times 100$ Gb/s RoCE network, and each node contains 4 Nvidia GPUs (40GB, without NVLink). 
\textit{b)} Cluster-B is equipped with 4 nodes connected with $4\times 200$ Gb/s InfiniBand network, and each node contains 8 Nvidia GPUs (80GB, with NVLink).

\MyPara{Models and Datasets.} 
We evaluate \ours{} on GPT3 models.
We carefully tune the hyper-parameters to saturate the GPU resources: 
\textit{1)} the number of transformer layers is set to be divisible by the PP dimensions before and after migration; 
\textit{2)} the hidden size is set to be divisible by TP dimensions before and after migration.
We train the models using OpenWebText dataset.

\MyPara{Baselines.}
We compare \ours{} with four checkpoint-based approaches (i.e., Megatron, Megatron-Dist~\cite{megatron-dist-checkpoint}, EasyCkpt~\cite{easyckpt}, and Fastest--their fastest combination) and a dynamic state management library (i.e., Tenplex~\cite{2025-sosp-tenplex}). 
\begin{enumerate}[label=\textit{\arabic*)}, leftmargin=5mm, itemsep=-0.25em, topsep=-0.25em]
\item Megatron -- the default checkpointing approach. As for reconfiguration, an extra convert script is used to reshard the checkpoint to another hybrid-parallel configuration.
\item Megatron-Dist -- an improved version of \textit{1)}. It enables distributed asynchronous saving and loading of the checkpoint. Moreover, it performs state resharding when loading the checkpoint, and the convert script in \textit{1)} is omitted. 
\item EasyCkpt -- an on-premise in-memory checkpointing. It caches the state in host memory and persists it to disk after many training steps. The convert script in \textit{1)} is required.
\item Fastest-baseline (FB) -- To represent the fastest checkpoint-based baseline, we pick the least times of \texttt{reshard}, and \texttt{save}/\texttt{load} phases from above baselines.  
\item Tenplex -- a state management library designed for LLM training. It supports dynamic parallelism transformation through an in-memory file system.
\end{enumerate}

\begin{table*}[t]
  \centering
  \footnotesize
  \vspace{-0.1in}
  \caption{Performance comparison of \ours{} with the checkpoint-based approaches.}
    \renewcommand{\arraystretch}{0.975}
    \begin{tabular}{l|l|l|c|c|c|c}
    \toprule
    \textbf{Reconfiguration} & \textbf{Method} & \textbf{Mem Type} & 
    \textbf{Save Ckpt (ms)} & 
    \textbf{Reshard (ms)} & 
    \textbf{Load Ckpt (ms)} & 
    \textbf{Total Time (ms)} \\
    \hline

    \multirow{5}{*}{\makecell[c]{Prev: PP8TP1\\ Curr: PP16TP1}}
    & Megatron & Disk & 11270.02 & \multirow{2}{*}{126405.59} & \multirow{2}{*}{5603.03} & 143278.63 \\
    \cline{2-4} \cline{7-7}
    & EasyCkpt & CPU Mem & 719.50 &  &  & 132728.12 \\
    \cline{2-7}
    & Megatron-Dist & Disk & 3267.21 & \multicolumn{2}{c|}{6185.45} & 9452.66 \\
    \cline{2-7}
    & Fastest Baseline & CPU Mem & 719.50 & \multicolumn{2}{c|}{3637.74} & 4357.24 \\
    \cline{2-7}
    \rowcolor{lightgray}
    \cellcolor{white} & \textbf{ETC (ours)} & \textbf{GPU Mem} & \textbf{\textbackslash} & \textbf{1500.88} & \textbf{\textbackslash} & \textbf{1500.88} \\
    \hline

    \multirow{5}{*}{\makecell[c]{Prev: PP4TP2\\ Curr: PP4TP4}} 
    & Megatron & Disk & 10347.99 & \multirow{2}{*}{228887.45} & \multirow{2}{*}{6837.16} & 246072.60 \\
    \cline{2-4} \cline{7-7}
    & EasyCkpt & CPU Mem & 961.69 &  &  & 236686.30 \\
    \cline{2-7}
    & Megatron-Dist & Disk & 1406.73 & \multicolumn{2}{c|}{7925.20} & 9331.93 \\
    \cline{2-7}
    & Fastest Baseline & CPU Mem & 961.69 & \multicolumn{2}{c|}{7480.16} & 8441.85 \\
    \cline{2-7}
    \rowcolor{lightgray}
    \cellcolor{white}& \textbf{ETC (ours)} & \textbf{GPU Mem} & \textbf{\textbackslash} & \textbf{1464.20} & \textbf{\textbackslash} & \textbf{1464.20} \\
    \hline

    \multirow{5}{*}{\makecell[c]{Prev: PP4TP4\\ Curr: PP4TP2}} 
    & Megatron & Disk & 10939.93 & \multirow{2}{*}{438597.94} & \multirow{2}{*}{5551.02} & 455088.89 \\
    \cline{2-4} \cline{7-7}
    & EasyCkpt & CPU Mem & 1036.85 &  &  & 445185.81 \\
    \cline{2-7}
    & Megatron-Dist & Disk & 1816.75 & \multicolumn{2}{c|}{8320.76} & 10137.51 \\
    \cline{2-7}
    & Fastest Baseline & CPU Mem & 1036.85 & \multicolumn{2}{c|}{7540.86} & 8577.71 \\
    \cline{2-7}
    \rowcolor{lightgray}
    \cellcolor{white} & \textbf{ETC (ours)} & \textbf{GPU Mem} & \textbf{\textbackslash} & \textbf{1347.83} & \textbf{\textbackslash} & \textbf{1347.83} \\
    \hline

    \multirow{5}{*}{\makecell[c]{Prev: PP4TP2\\ Curr: PP6TP4}} 
    & Megatron & Disk & 13791.80 & \multirow{2}{*}{241625.13} & \multirow{2}{*}{7228.64} & 262645.57 \\
    \cline{2-4} \cline{7-7}
    & EasyCkpt & CPU Mem & 1034.46 &  &  & 249888.23 \\
    \cline{2-7}
    & Megatron-Dist & Disk & 1662.15 & \multicolumn{2}{c|}{9882.39} & 11544.54 \\
    \cline{2-7}
    & Fastest Baseline & CPU Mem & 1034.46 & \multicolumn{2}{c|}{9254.7} & 10289.16 \\
    \cline{2-7}
    \rowcolor{lightgray}
    \cellcolor{white} & \textbf{ETC (ours)} & \textbf{GPU Mem} & \textbf{\textbackslash} & \textbf{4425.39} & \textbf{\textbackslash} & \textbf{4425.39} \\

    \hline
    \multirow{5}{*}{\makecell[c]{Prev: PP6TP4\\ Curr: PP4TP2}} 
    & Megatron & Disk & 9127.24 & \multirow{2}{*}{481885.34} & \multirow{2}{*}{6943.09} & 497955.67 \\
    \cline{2-4} \cline{7-7}
    & EasyCkpt & CPU Mem & 962.63 &  &  & 489791.06 \\
    \cline{2-7}
    & Megatron-Dist & Disk & 1210.80 & \multicolumn{2}{c|}{8117.31} & 9328.11 \\
    \cline{2-7}
    & Fastest Baseline & CPU Mem & 962.63 & \multicolumn{2}{c|}{7869.14} & 8831.77 \\
    \cline{2-7}
    \rowcolor{lightgray}
    \cellcolor{white} & \textbf{ETC (ours)} & \textbf{GPU Mem} & \textbf{\textbackslash} & \textbf{3577.33} & \textbf{\textbackslash} & \textbf{3577.33} \\

    \hline
    \multirow{5}{*}{\makecell[c]{Prev: PP4TP2DP3\\ Curr: PP4TP8DP1}} 
    & Megatron & Disk & 11298.80 & \multirow{2}{*}{231999.41} & \multirow{2}{*}{9414.73} & 252712.94 \\
    \cline{2-4} \cline{7-7}
    & EasyCkpt & CPU Mem & 1475.39 &  &  & 243516.26 \\
    \cline{2-7}
    & Megatron-Dist & Disk & 1747.25 & \multicolumn{2}{c|}{11241.13} & 12988.38 \\
    \cline{2-7}
    & Fastest Baseline & CPU Mem & 1475.39 & \multicolumn{2}{c|}{10969.26} & 12444.66 \\
    \cline{2-7}
    \rowcolor{lightgray}
    \cellcolor{white} & \textbf{ETC (ours)} & \textbf{GPU Mem} & \textbf{\textbackslash} & \textbf{4396.12
} & \textbf{\textbackslash} & \textbf{4396.12} \\

    \bottomrule
    \end{tabular}
  \label{tab:checkpoint_stats}
\end{table*}

\subsection{Comparison with Checkpoint-based Approaches}
Under various scenarios of hybrid-parallel reconfiguration, we evaluate the model state migration time of \ours{} and the checkpoint-based approaches. 
We train GPT3-32B model on Cluster-A. The hidden size is 3072, and the transformer layer number is 36. In the scenario of reconfiguring PP8TP1 to PP16TP1, since neither 8 nor 16 is divisible by 36, we set the transformer layer number to 32.
Experimental results are shown in Table~\ref{tab:checkpoint_stats}.
Since EasyCkpt does not optimize the \texttt{reshard} and \texttt{load} phases, we have merged EasyCkpt and Megatron in these columns.
Compared with checkpoint-based baselines
, \ours{} achieves superior performance across all reconfigurations. 

In the \texttt{save} phase, Megatron-Dist and EasyCkpt achieve 6.52$\times$ and 11.24$\times$ on average compared to Megatron, because Megatron-Dist adopts distributed I/O to save, and EasyCkpt just offloads the checkpoint to host memory.  
In the \texttt{reshard} and \texttt{load} phase, Megatron-Dist shows 21.34$\times$--60.22$\times$ speedup compared to Megatron and EasyCkpt. This is because it fuses the two phases to eliminate redundant I/O, and it adopts distributed I/O to load the checkpoint to host memory. 
With "Fastest-baseline (FB)", we combine EasyCkpt's advantage in \texttt{save} and Megatron-Dist's advantage in \texttt{reshard} and \texttt{load}. 
However, \ours{} still achieves 2.33$\times$--6.37$\times$ speedup than FB.
This is because: 
\textit{1)} with \ours{}, the model state is completed within the GPU memory, the checkpoint persistence and offloading is completely eliminated.
\textit{2)} \ours{} directly sends and receives the model state with inter-GPU communication, which usually has higher bandwidth than offloading and I/O. 
\textit{3)} the model state resharding of \ours{} is inherently parallel and leverages locality to minimize the data movement.

\subsection{Comparison with the State Management Library Tenplex}
We compare \ours{} with Tenplex on Cluster-B. 
For fair comparison, the model configurations (i.e., GPT3-2.7B and GPT3-6.7B) and the scenarios of hybrid-parallel reconfiguration are consistent with the open-source repository of Tenplex. 
The reconfigurations involve the scaling in TP, PP, and DP. 
We compare the Tenplex and Tenplex-Central modes, where Tenplex has distributed state management between State Transformer instances on different nodes and Tenplex-Central performs all model state resharding at the central node.

\begin{figure}[htbp]
    \centering
    \includegraphics[width=\linewidth]{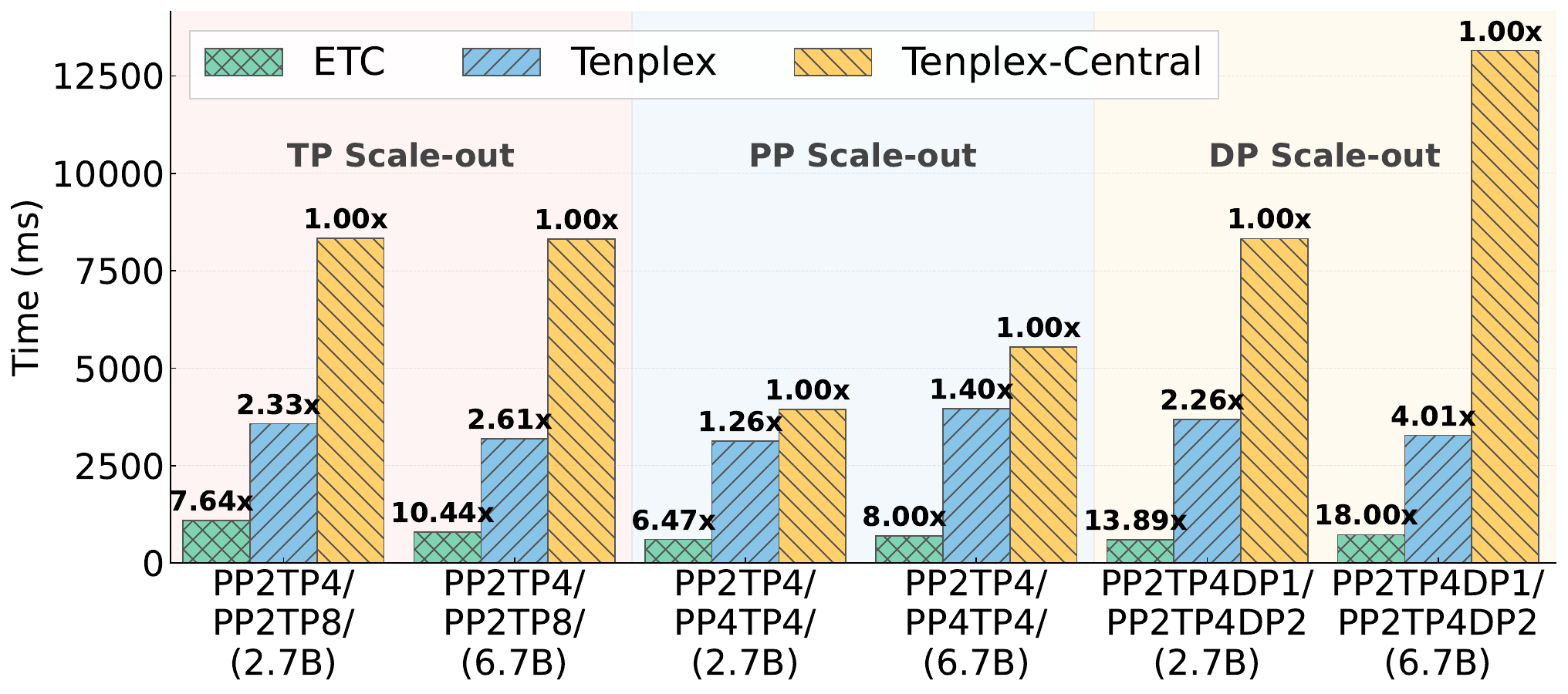}
    \caption{Comparison between ETC and Tenplex. Y-axis represents the time(ms) of state migration, and x-axis represents different reconfiguration scenarios.
    }
    \label{fig:ETC_vs_Tenplex}
\end{figure}

As shown in Figure~\ref{fig:ETC_vs_Tenplex}, \ours{} achieves 4.79$\times$ and 10.74$\times$ speedup compared to Tenplex and Tenplex-Central, respectively. 
This is attributed to two reasons. First, \ours{} maintains the cost matrix to represent the data movement and solves the cost matrix to generate the migration instructions with minimal data movement overhead. Second, \ours{} transfers the model state from GPU to GPU directly, while Tenplex still adopts a checkpoint-based design.

\subsection{Benefits of Communication Coalescing}
\label{sec:comm_fusion}

We evaluate the benefits brought by coalescing the communication in migration and de-fragmentation stages. 
We adopt the hybrid-parallel reconfigurations, including \textit{1)} scale in PP, \textit{2)} scale out TP, and \textit{3)} scale them simultaneously.
We also set different transformer layers ($l$) and hidden sizes ($h$).

\begin{figure}[h]
    \centering
    \includegraphics[width=\linewidth]{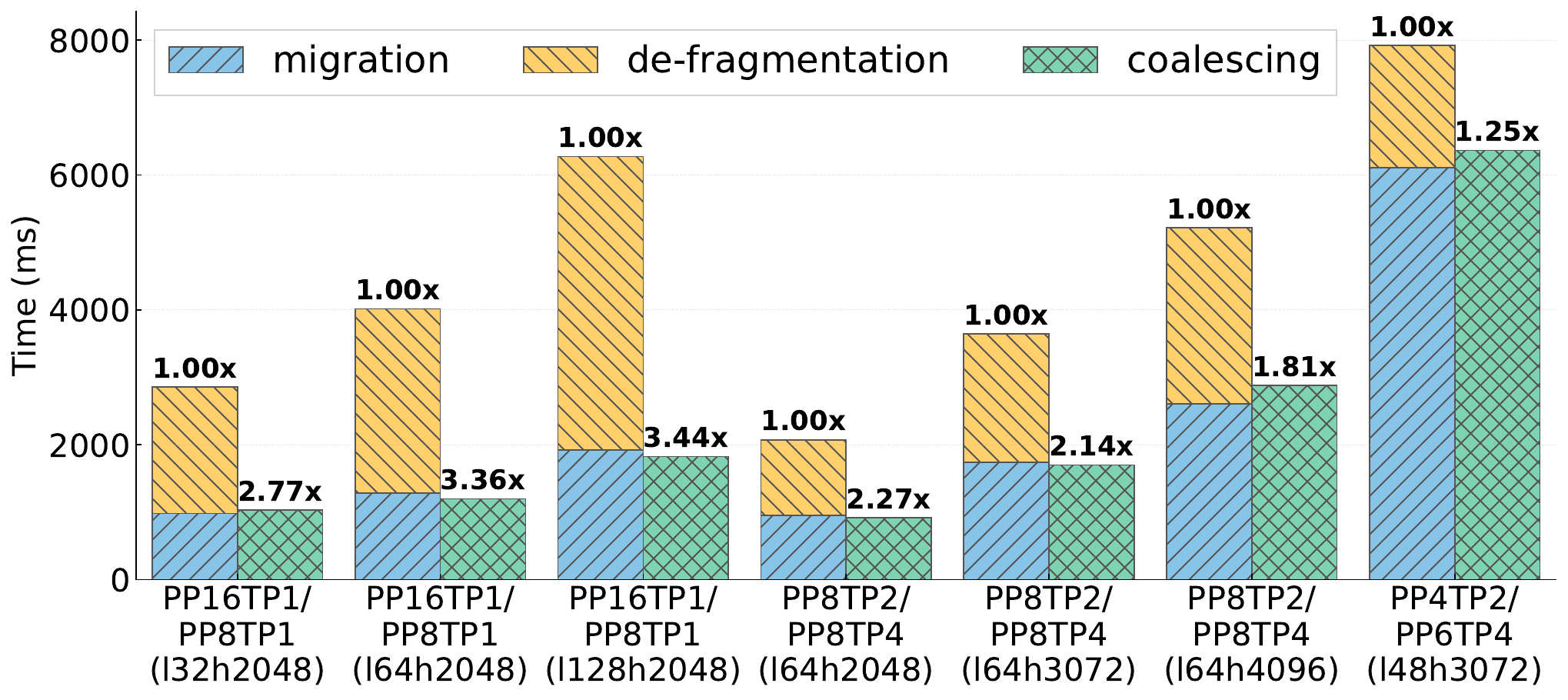}
    \caption{Communication time of migration instructions, de-fragmentation instructions, and the coalesced instructions.}
    \label{fig:Comm_Fusion}
\end{figure}

As shown in Figure~\ref{fig:Comm_Fusion}, the coalesced communication achieves 2.43$\times$ speedup on average compared to the two-stage communication.
In the first three scenarios, we keep the same parallel configuration and increase the transformer layer number from 32 to 128. With the total parameters increase, the time required for two stages is on the rise. 
However, the communication time after coalescing is almost the same as the time of state migration. 
This is because the primary ability of communication coalescing is to change the destination to which the workload is sent. 
According to our observation, the communication coalescing does not alter the maximum communication volume across devices throughout the whole hybrid-parallel reconfiguration.

\subsection{Performance Improvement After De-fragmentation}

\begin{figure}[h]
    \centering
    \includegraphics[width=\linewidth]{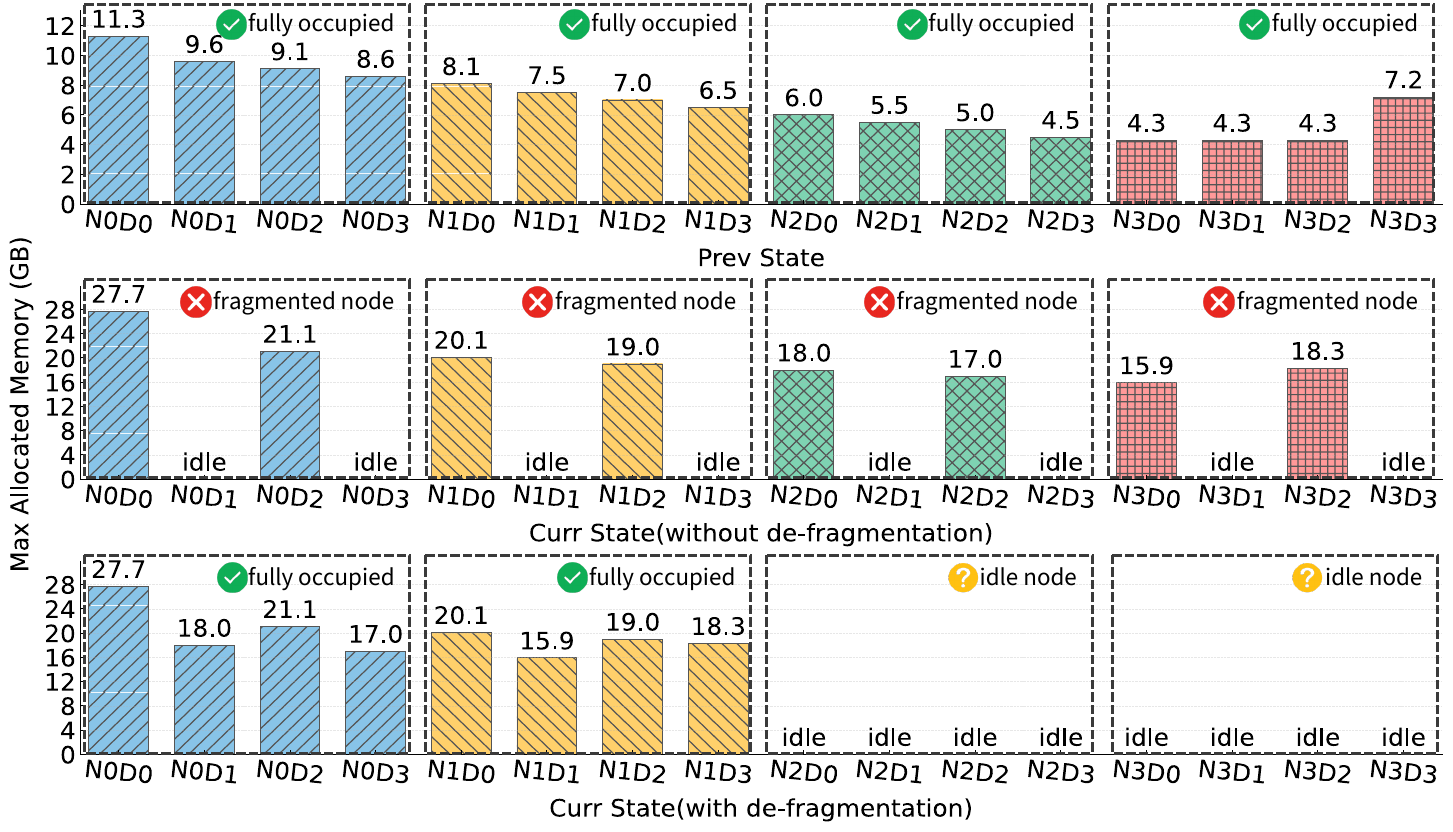}
    \caption{Memory usage per GPU when changing from PP16TP1 to PP8TP1.
    Y-axis represents the maximum allocated memory, and x-axis represents different GPUs. The labels on the x-axis are in the format of NxDy, indicating the $y$-th device in node $x$.
    }
    \label{fig:Node_Fragement_Mem}
\end{figure}

To demonstrate the utilization after resource de-fragmentation, we design an experiment on Cluster-A, where each node has 4 GPUs.
The training job of GPT3-64B changes from PP16TP1 to PP8TP1, where the number of GPUs is reduced by half.
As shown in Figure~\ref{fig:Node_Fragement_Mem}, before de-fragmentation, 2 GPUs in each node are occupied and the others are idle, making it impossible to release resources on a node-by-node basis. After the node de-fragmentation of \ours{}, the occupied GPUs are concentrated in two nodes.

To demonstrate the performance improvement after de-fragmentation, we design an experiment on Cluster-B. 
The training job of GPT3-64B changes from PP4TP2 to PP4TP4.
From PP4TP2 to PP4TP4 reconfiguration, the number of GPUs doubles. However, before de-fragmentation, the per-iteration time of the model increases. After de-fragmentation of \ours{}, the per-iteration time decreases by 26.7\% by leveraging the NVLink in TP communication.

\subsection{Scalability}
To demonstrate the scalability of \ours{}, we analyze time breakdown of \ours{} under various hybrid-parallel reconfigurations on Cluster-A. 
We divide the entire elastic migration into 5 parts: 
\textit{1)} \texttt{cost matrix}, which includes the generation of cost matrix and the coarse-grained migration sketch; 
\textit{2)} \texttt{execution}, which performs the fine-grained migration instructions through the execution engine; 
\textit{3)} \texttt{build group}, which represents reconstructing a global communication group with re-assigned devices; 
and \textit{4)} \texttt{others}, which includes the other modules, such as the meta model. 
Among them, \textit{1)2)4)} belong to the state migration of \ours{}. 
And \texttt{build group} relies on the training framework implementation (e.g., PyTorch and Megatron) and the hardware configuration (e.g., network and GPU topology), and its time varies significantly. 
For example, on Cluster-B, the time to construct a communication group with 4 nodes (32 GPUs) is around 1.5s, while on Cluster-A, the time with the 32 GPUs is around 3.0s.

\begin{figure}[htbp]
    \centering
    \includegraphics[width=\linewidth]{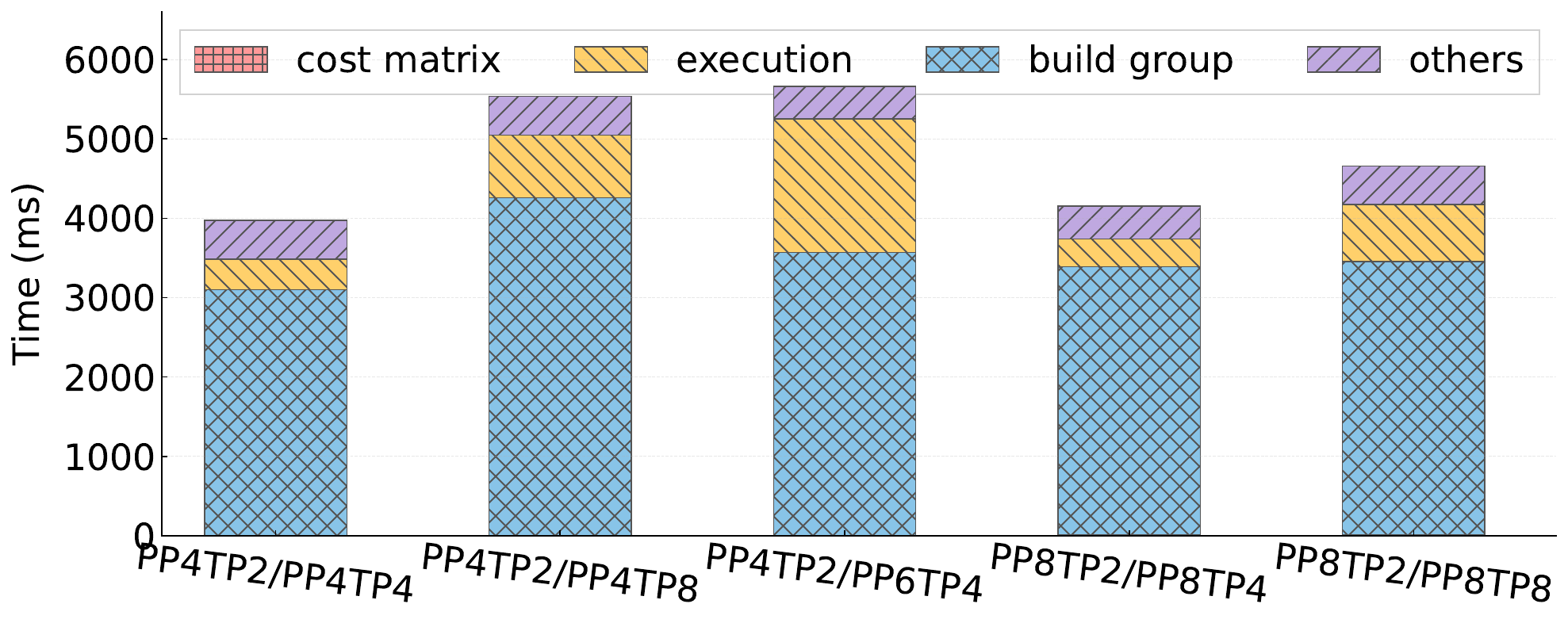}
    \caption{Time breakdown of ETC. }
    \vspace{-0.1in}
    \label{fig:breakdown}
\end{figure}

As shown in Figure~\ref{fig:breakdown}, we perform experiments where PP $\times$ TP ranges from 8 to 64, which covers the common ranges in Megatron-LM repository. 
It is obvious that the time of \texttt{cost matrix} is almost negligible (7.54ms on average). 
And \texttt{execution} accounts for a larger proportion of time, because of communication. 
The time of \texttt{execution} is affected by three factors: \textit{1)} the size of the model parameters; \textit{2)} the network topology; and \textit{3)} the complexity of migration sketch. 
Factors \textit{1)} and \textit{2)} are quite intuitive, and factor \textit{3)} explains why the \texttt{execution} time for PP4TP2/PP6TP4 is slightly longer. 
In this migration, both PP size and TP size are changed. It leads to a more complex migration sketch, where some devices need to sequentially execute \texttt{Recv} and \texttt{Send} instructions.

\subsection{Memory Footprint and Accuracy Analysis}
\label{sec:mem_accuracy}

We conduct a case study on a GPT3-6.7B model. The initial configuration is PP8TP1 on 2 nodes of Cluster-A. At iteration 20, elastic migration is triggered to expand the training job to 4 nodes with a configuration PP16TP1. We profile the GPU memory allocation on rank 0 throughout the migration process and record the training loss across 100 iterations.

\begin{figure}[h]
    \centering
    \includegraphics[width=\linewidth]{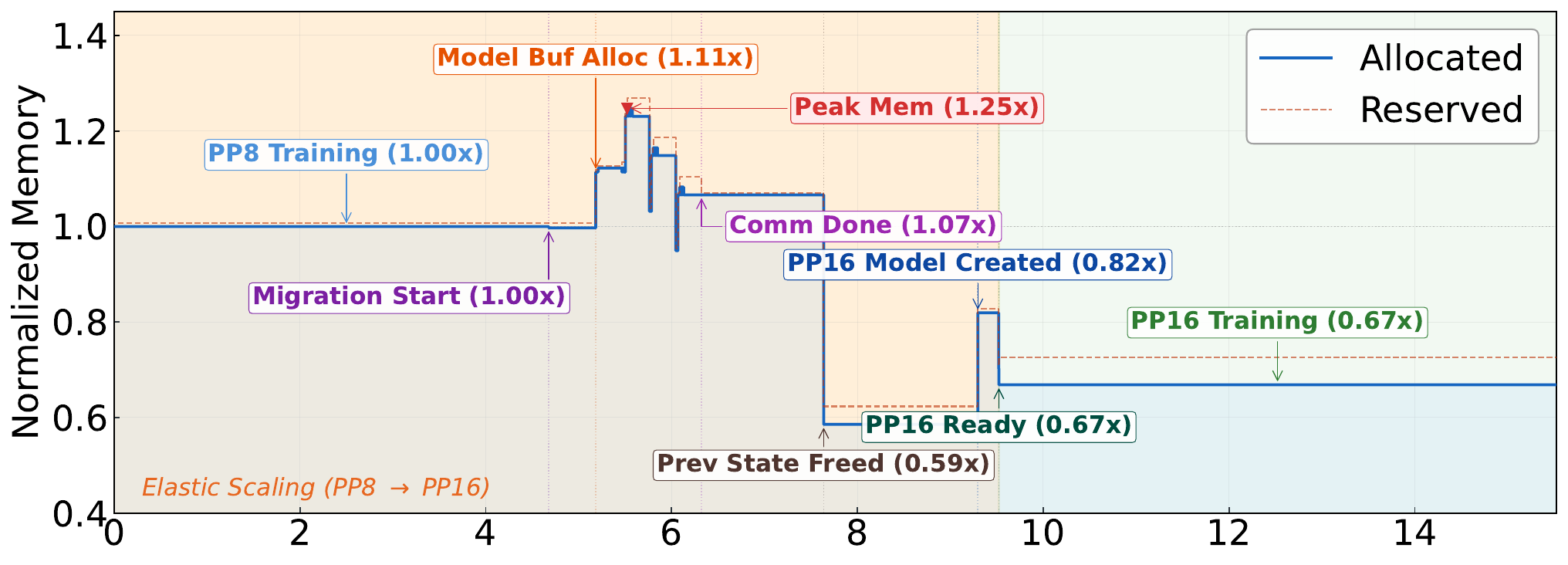}
    \vspace{-0.1in}
    \caption{Memory footprint under elastic migration (PP8$\to$PP16). }
    \label{fig:memory_timeline}
\end{figure}

Figure~\ref{fig:memory_timeline} shows the normalized GPU memory during migration, with all values normalized to the PP8 steady-state. 
The memory footprint exhibits three distinct phases. 
\textit{1)~Peak phase:} Upon triggering migration, communication buffers are allocated for receiving resharded model parameters (1.11$\times$). As optimizer states are migrated via \ours{} in a pipelined manner, the co-residence of previous and current states causes a transient peak of 1.25$\times$. \ours{} reduces the peak to one optimizer component at a time, avoiding doubling all states. 
\textit{2)~Cleanup phase:} When communication completes, the previous PP8 states are freed (0.59$\times$), and the new PP16 model/optimizer is reconstructed (0.82$\times$). 
\textit{(3)~PP16 steady state:} The final memory is 33\% lower than PP8, as doubling the pipeline depth halves the per-rank layer count.

\begin{figure}[h]
    \centering
    \includegraphics[width=\linewidth]{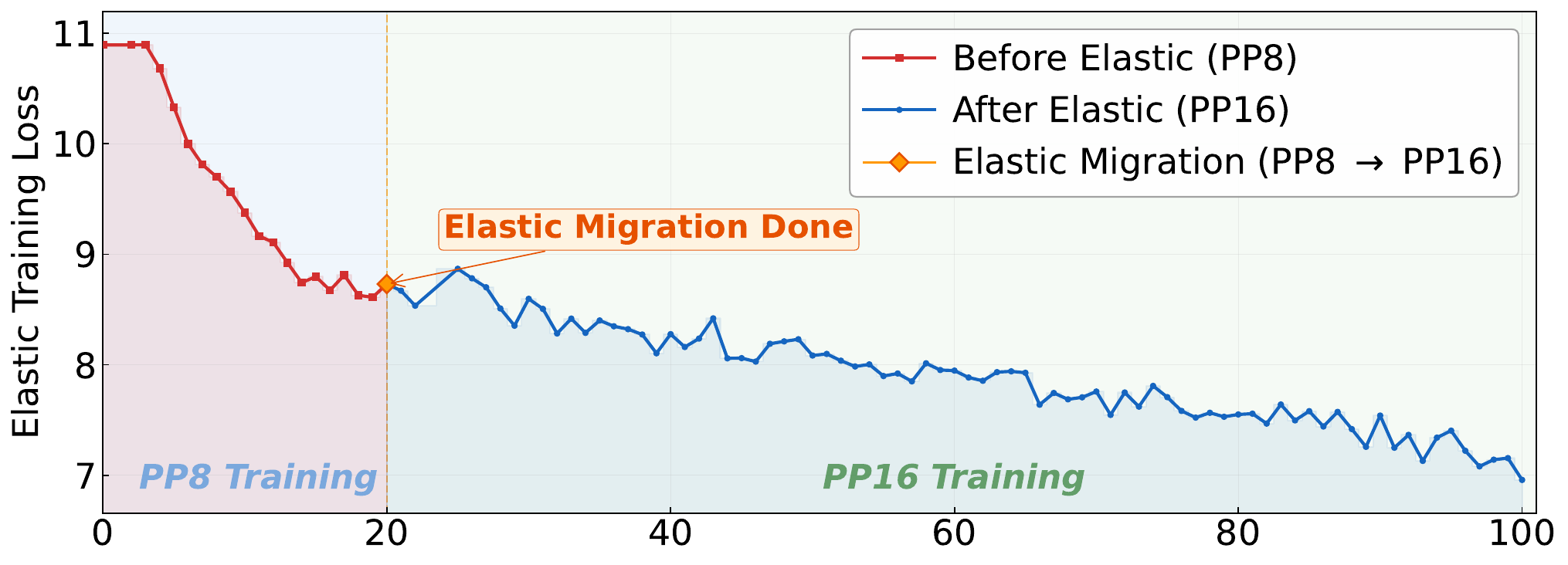}
    \vspace{-0.2in}
    \caption{Loss curve. Elastic migration happens at 20-th iteration (PP8$\to$PP16). }
    \vspace{-0.2in}
    \label{fig:loss_curve}
\end{figure}

Figure~\ref{fig:loss_curve} presents the training loss over 100 iterations with elastic migration at iteration~20. 
Two observations confirm the correctness of our approach. First, the loss curve shows no spike or discontinuity at the migration boundary, the loss transitions seamlessly from PP8 to PP16, Second, after transitioning to PP16, the loss continues to decrease, following the same convergence trend. By ensuring that all optimizer states and model parameters are transferred with bit-exact fidelity, and keeping the global batch size and learning rate consistent during migration, \ours{} achieves elastic migration without any significant impact on loss.

\section{Related Works}
\label{sec:relatedworks}

\MyPara{Elastic Training.}
It allows a job to continue its training procedure under varying resources.
TorchElastic~\cite{pytorch_elastic} and Horovod Elastic~\cite{horovod_elastic} support elastic training and fault tolerance mainly on the number of participating worker processes (e.g., one worker per GPU), and thus their native implementation only supports elasticity on data parallel. 
Based on them, hybrid-parallel training frameworks, such as Megatron-LM~\cite{2019-arxiv-megatron-lm} and DeepSpeed~\cite{2022-kdd-deepspeed}, can also support elastic training, but need developers to tackle the job states with either checkpoint-based approach (e.g., PcCheck~\cite{2025-asplos-pccheck}, ByteCheckpoint~\cite{2025-nsdi-bytecheckpoint}) or checkpoint-free migration approach (e.g., this paper).
KungFu~\cite{2020-osdi-kungfu} and Pollux~\cite{2021-osdi-pollux} support adjusting the hyper-parameters (e.g., batch sizes) of elastic training jobs to enable various scheduling policies.
EasyScale~\cite{2023-sc-easyscale} enables efficient and accuracy-consistent elastic training but only focuses on data parallel. 
Varuna~\cite{2022-eurosys-varuna} allows elasticity on pipeline dimension while requiring users to specify the cutting points on models.
Oobleck~\cite{2023-sosp-oobleck} focuses on fault tolerance but provides limited elasticity by pre-defined pipeline templates. 
In contrast, \ours{} focuses on the time-consuming state migration of hybrid-parallel reconfiguration, which is frequent in the elastic training scenario. 

\MyPara{Auto Parallelization.}
With the given resources and target models, searching for the optimal hybrid-parallel configuration is difficult.
Alpa~\cite{2022-osdi-alpa} partitions the compute cluster into meshes, assigns model stages to meshes, and orchestrates the inter- and intra-operator operation passes. 
Unity~\cite{2022-osdi-unity} represents the parallel optimizations and algebraical transformations as graph substitutions in the computation graph of distributed training, and adopts a hierarchical search algorithm.
And \ours{} can help the training job to efficiently migrate to the searched hybrid-parallel configurations.

\section{Conclusion}
\label{sec:conclusion}

With \ours{}, we demonstrate the success of checkpoint-free model state migration for elastic hybrid-parallel LLM training.  
Compared to the checkpoint-based migrations, \ours{} performs inter-GPU communications without any I/O persistence. 
We try the best to keep model state in place, exploit a unique cost matrix to minimize the data movement during migration, and avoid resource fragmentation through communication coalescing, enabling second-level migrations especially on clusters co-locating inference and training jobs.
In the future, we hope that \ours{} will make the elastic training more practical in production environments for cluster utilization.

\bibliographystyle{IEEEtran}
\bibliography{references.bib}

\end{document}